\documentclass[11pt,preprint]{aastex}
\usepackage{amsmath,amssymb}
\usepackage{xspace}
\usepackage{graphicx}
\usepackage{epstopdf}  
\usepackage{graphicx}
\usepackage{epsfig}
\usepackage{verbatim}
\usepackage{morefloats} 
\usepackage[colorlinks,urlcolor=blue,citecolor=black,linkcolor=blue]{hyperref}
\usepackage{CJK} 

\def\beq{\begin{equation}}
\def\eeq{\end{equation}}

\def\mps{m~s$^{-1}$}
\def\msun{M_{\odot}}

\def\leq{\leqslant}
\def\geq{\geqslant}
\def\micron{$\mu$m}
\def\percm2{{\rm cm^{-2}}}

\def\kepler{\textit{Kepler}}
\def\spitzer{\textit{Spitzer}}
\def\micron{$\mu$m}
\def\mjup{$M_{\rm Jup}$}

\def\rearth{R_\oplus}
\def\mearth{M_\oplus}

\slugcomment{}
\shorttitle{}
\shortauthors{}

\begin{document}

\begin{CJK*}{UTF8}{gbsn}

\title{The Exoplanet Orbit Database \textsc{II}: Updates to
  exoplanets.org}

\author{Eunkyu Han\altaffilmark{1,2}}
\author{Sharon X.\ Wang\altaffilmark{1,2,3}}
\author{Jason T.\ Wright\altaffilmark{1,2,3}}
\author{Y.\ Katherina Feng\altaffilmark{1,2}}
\author{Ming Zhao\altaffilmark{1,2}}
\author{Onsi Fakhouri\altaffilmark{4}}
\author{Jacob I.\ Brown\altaffilmark{1,2}}
\author{Colin Hancock\altaffilmark{1,2}}

\altaffiltext{1}{Department of Astronomy \& Astrophysics, 525 Davey
  Lab, The Pennsylvania State University, University Park, PA 16802, USA;
  Send correspondence to jtwright@astro.psu.edu.}
\altaffiltext{2}{Center for Exoplanets and Habitable Worlds, The
  Pennsylvania State University, University Park, PA 16802, USA}
\altaffiltext{3}{Penn State Astrobiology Research Center, The
  Pennsylvania State University, University Park, PA 16802, USA} 
\altaffiltext{4}{Pivotal Labs, Cloud Foundry, 875 Howard St., Fourth
  Floor, San Francisco, CA 94103}

\begin{abstract}

The Exoplanet Orbit Database (EOD) compiles orbital, transit, host
star, and other parameters of robustly detected exoplanets reported in
the peer-reviewed literature. The EOD can be navigated through the
Exoplanet Data Explorer (EDE) Plotter and Table, available on the
World Wide Web at exoplanets.org. The EOD contains data for 1492
confirmed exoplanets as of July 2014.  The EOD descends from a table
in \citet{Butler2002} and the Catalog of Nearby Exoplanets
\citep{Butler2006}, and the first complete documentation for the EOD
and the EDE was presented in \cite{Wright2011}. In this work, we
describe our work since then.  We have expanded the scope of the EOD
to include secondary eclipse parameters, asymmetric uncertainties, and
expanded the EDE to include the sample of over 3000 \kepler\ Objects
of Interest (KOIs), and other real planets without good orbital
parameters (such as many of those detected by microlensing and
imaging). Users can download the latest version of the entire EOD as a
single comma separated value file from the front page of
exoplanets.org.

\end{abstract}  

\section{Introduction}\label{sec:intro}

Since the launch of NASA's \kepler\ mission \citep{Borucki2010}, the
number of confirmed exoplanets has increased from about 300 to over
1490 as of July 2014. \kepler\ contributed to a majority of these new
discoveries ($> 800$) while providing an additional sample of over
3,000 exoplanet candidates (\kepler\ Objects of Interest, KOIs;
e.g.~\citealt{Batalha2013}). The transit method has thus surpassed the
radial velocity (RV) method to become the most fruitful means of
detecting exoplanets. Meanwhile, great promise lies within the new and
future exoplanet surveys and instruments with direct imaging (e.g.,
the Gemini Planet Imager, GPI, \citealt{Macintosh2014}) and
microlensing (e.g., the Wide-Field InfraRed Survey Telescope, WFIRST,
\citealt{Green2012}), while the fronts of RV and transit exoplanet
searches keep expanding (e.g., the MINiature Exoplanet Radial Velocity
Array, MINERVA, \citealt{Wright2014}; the Habitable-zone Planet Finder,
HPF, \citealt{Mahadevan2012} and the Transiting Exoplanet Survey
Satellit, TESS, \citealt{Ricker2014}).

This paper describes our continuing efforts since \cite{Butler2006}
and \cite{Wright2011} to maintain the Exoplanet Orbit
Database\footnote{See \url{http://expoplanets.org}}(EOD) and the
Exoplanets Data Explorer (EDE), to keep track of exoplanet
discoveries, and to better catalog orbital parameters and host star
properties.

There are other entities devoted to similar efforts,
including the Extrasolar Planets Encyclopaedia\footnote{See
  \url{http://exoplanet.eu}.}  \citep{Schneider2011} and the NASA
Exoplanet Archive\footnote{See
  \url{http://exoplanetarchive.ipac.caltech.edu}.}
\citep{Akeson2013}.  The natures of these efforts are largely
complementary to the EOD, and to the
degree that they are redundant provide important cross-checking and
validation.  As described in \cite{Wright2011}, the EOD is
dedicated to cataloging properties of all \emph{robustly-detected} exoplanets with
\emph{well-determined} orbital parameters published in
\emph{peer-reviewed} literature, and providing outstanding data
exploration and visualization tools for these planets.

The EOD is hosted at the Web domain exoplanets.org, which also hosts
the EDE, an interactive table with plotting tools for all the EOD
exoplanets, in addition to other confirmed exoplanets without good
orbital parameters, and KOIs (except for those with a ``FALSE
POSITIVE'' disposition at the Exoplanet Archive). The entire dataset
of exoplanets.org is available for download as a single comma
separated value file from the Website front page.\footnote{At
  \url{http://exoplanets.org/csv-files/exoplanets.csv}.}  An example
of this CSV file is available with the electronic version of this
paper.

The EOD is widely
used in exoplanet research (e.g., \citealt{Dawson2013},
\citealt{Howard2013}, \citealt{Kipping2013}, \citealt{Kane2014},
\citealt{Weiss2014}), and the website exoplanets.org had over $47,000$ visits in
2013\footnote{Measured as ``visits" by Google Analytics}.

Since the documentation described in \cite{Wright2011}, the EOD has expanded its scope: we now do
not require the host star of an exoplanet to be a normal star (e.g.,
exoplanets around pulsars are now included). We now ingest more
parameters describing the exoplanet systems and their host stars, and some
parameters have been revised or deleted. We describe our updates to the
EOD in Section~\ref{sec:update}. The EDE at exoplanets.org has also
expanded its scope and now includes the \kepler\ exoplanet candidates
(KOIs). We describe these changes in Section~\ref{sec:kepler}. The
website exoplanets.org has gone through major back-end changes since
\cite{Wright2011}, which are described in Section~\ref{sec:website}. We
summarize the updates and possible future improvements in
Section~\ref{sec:summary}.

\section{Updates on the EOD}\label{sec:update}

The scope of the EOD has expanded, and there are three major changes
from what was described in \cite{Wright2011}. Previously, we only
included exoplanets orbiting the ``normal stars", but now we are not
limiting the type of host stars (e.g. we included exoplanets orbiting
neutron stars). Furthermore, in \cite{Wright2011}, we only included
exoplanets that were discovered by the RV and transit methods in the
EOD, but now we also include planets discovered via timing (including
pulsar planets and those discovered by eclipse timing). We now include
exoplanets discovered by any other methods with robust detections and
good orbital parameters, such as $\beta$ Pictoris $b$
\citep{Chauvin2012,Nielsen2014,Macintosh2014}.  

Previously, we used a generous mass cutoff, recognizing that any
definition of ``exoplanet'' will not necessarily track with the
formation mechanism for an object (which may, after all, be unknown or
unknowable).  Our previous mass cutoff was a minimum mass (m$\sin{i}$)
smaller than 24\mjup; to make this cutoff more relevant to a wide
range of stellar masses and detection methods, we now require the measured
planet-star mass ratio to be smaller than 24 \mjup$/\msun$ ($< 0.023)$.

The rest of the selection criteria for the EOD remain unchanged. We
still require the planets to have robust detections and
well-determined orbits (though these are not strictly quantitative and
sometimes subjective; see Section~2 of \citealt{Wright2011}). If new
evidence comes to light and we are no longer confident about a planet,
we may remove it from EOD, even if the new evidence is not peer
reviewed.  The rationale for this scheme is, as it has been since the
CNE, that the EOD represents our best, expert effort to produce a
clean set of trustworthy parameters.  We do not report \textit{all}
the announced exoplanets as we are not trying to provide ``an
encyclopedic presentation of every claimed detection of an exoplanet"
\citep{Wright2011}.  More comprehensive sets of planet parameters are
available from other exoplanet compendia on the Web.

\begin{figure}[!htb]
\centering
\includegraphics[width=\textwidth]{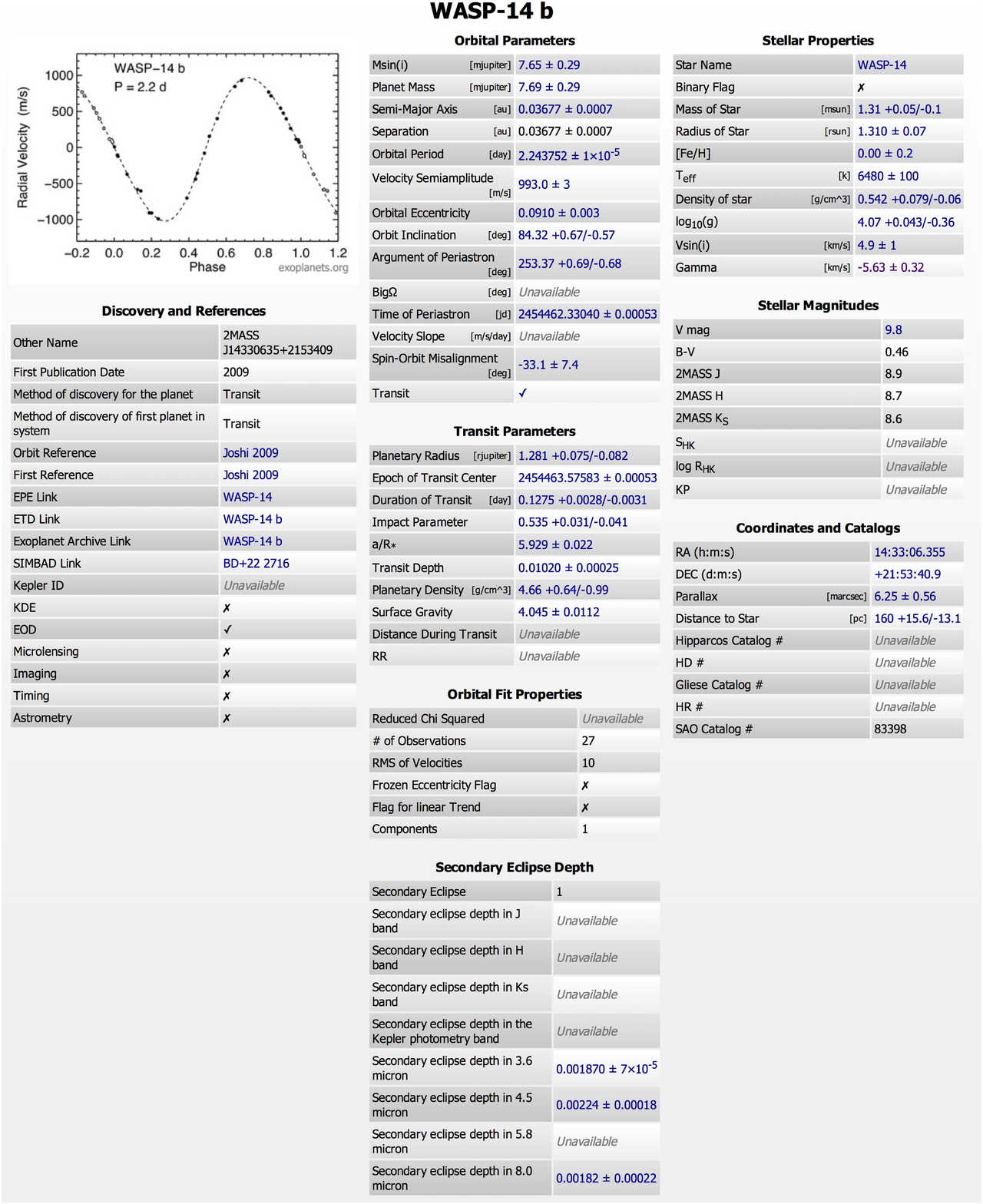}
\caption{An example page for an individual exoplanet (WASP-14b in this
  case). The numbers listed in each field contain links to the
  reference on NASA ADS. In this example for WASP-14b, we provide the plot
  for the phased RVs, but we have yet to implement such feature for all
  planets with available RV data.}
\label{fig:individual}
\end{figure}

In the following subsections, we describe the newly added and revised
fields since \cite{Wright2011}, and list all fields in
Table~\ref{tab:par}. We divide the fields into categories consistent
with the individual planet page on
exoplanets.org. (Figure~\ref{fig:individual} is an example of
individual planet page, which is presented to website users who click
on a planet's name). Some fields are not explicitly tabulated on the
page (e.g., the URLs for ADS pages for papers providing the underlying
data), but are explicitly listed in the downloadable {\tt .csv} file
(see Section~\ref{sec:intro}). Though we focus on the EOD in this
section, the fields apply to all planets on exoplanets.org. We follow
the definition of orbital elements in \cite{Wright2013} unless
indicated otherwise.

\subsection{Discovery and References}\label{sec:disc}

We report general information of when, by whom and how the planets
were discovered and first reported and provide the references for
underlying data. We describe the added or revised fields since
\cite{Wright2011} below.

The DISCMETH field previously reported the method of discovery,
which was either ``radial velocity'' or ``transit''. We have replaced DISCMETH
with {\bf PLANETDISCMETH} and {\bf STARDISCMETH} to better
report how exoplanet systems and individual exoplanets are first
discovered. PLANETDISCMETH reports the discovery method for the
planet, while STARDISCMETH reports the discovery method for the {\it
  planetary system} (i.e. the {\it first}
discovered planet around the host star). These fields are stored
in string format and have one of the following values: ``RV",
``Transit", ``Imaging", ``Microlensing", ``Timing", or ``Astrometry"
(no planets have a value of ``Astrometry" yet since no planet has yet
been {\it discovered} via astrometry).

The {\bf MONTH} field reports the integer month of the
peer-reviewed paper publication date.

{\bf KEPID} stands for KEPler IDentification, which is the integer number assigned
for a Kepler host star. KEPID is only available for unconfirmed
KOI's imported directly from the Exoplanet Archive by the
exoplanets.org backend. 

{\bf KOI} stands for Kepler Object of Interest, which is an object
with transit signals detected by \kepler\ that have not been validated
as being due to exoplanets.  The KOI field defined by the \kepler\
team and shown on the Exoplanet Archive is stored as a floating point
number, consisting of a whole number designated to the host star
followed by a decimal portion denoting a candidate planet. For example,
KOI 102 is a star suspected to host two exoplanets, and KOI 102.01 is
the inner planet candidate and KOI 102.02 is the outer one. Once a
candidate is confirmed, the {\bf NAME} field of the planet is replaced by
the official \kepler\ ID (``Kepler" followed by a hyphen, an integer,
a space, and finally a lower case letter). The KOI number is then used as
{\bf OTHERNAME}. For more information, see
Section~\ref{sec:kepler}.

{\bf EOD} is a Boolean which indicates whether an exoplanet is included
in the EOD for having well-characterized orbital
parameters, such as the orbital period (PER). We added this field because exoplanets.org
now also includes robustly discovered planets that do not have
well-characterized orbits, and also \kepler\ planet candidates (see
Section~\ref{sec:kepler} for more details).  Planets on exoplanets.org
and appearing in the EDE that are not part of the EOD have EOD set to
FALSE (i.e.\ zero).

{\bf MICROLENSING, IMAGING, TIMING} and {\bf ASTROMETRY} are Boolean flags which
indicate that a planet has been detected via these methods. For example,
MICROLENSING $=$ 1 means the planet is detected in a microlensing
event. Note that planets may be detected with multiple methods.

\subsection{Orbital Parameters}\label{sec:orbit}

We report the orbital parameters that determine the shape of the
planet's orbit. Depending on the detection methods, different sets of
parameters are given by the literature. For instance, if a planet is
detected by the radial velocity method, the semiamplitude, $K$, is
reported, whereas the inclination of the orbit, $i$, is usually only
known via the transit method (although planet-planet interactions and
astrometry can also determine it). We describe the added fields since
\cite{Wright2011} below.

{\bf MASS} is the mass of the planet in the unit of Jupiter
mass. Previously, we only reported the minimum mass, ($m\sin{i}$,
MSINI) of the planets.  We now, additionally, report the best estimate
of the true mass in a heterogeneous way.  For example, if a planet is discovered by microlensing, we
report the actual mass of the planet according to the literature. If a
planet has both radial velocity and transit measurements, we calculate
the mass from the minimum mass $m\sin{i}$ and the orbital inclination
$i$. For convenience of plotting RV and other planets on a common mass
scale (see Section~\ref{sec:kepler}), MASS is set to MSINI if the
orbital inclination is not known and the reference for MASS ({\bf
  MASSREF}) is set to ``Set to MSINI; I unknown''. 

Also for plotting convenience, for transiting $Kepler$ planets without
$m\sin{i}$ measured, we estimate their mass using mass-radius
relations from the literature. We follow a series of relations given
by

\[
M_p=
\begin{cases}
M_{\rm WM} &  R<\rearth \\
(1-x)M_{\rm WM} + xM_L  &  \rearth \leq R<4\rearth \\
M_L & 4 \rearth \leq R < 6 \rearth \\
\min{(M_{\rm M},{\text \mjup})} &  6\rearth \leq R < b \\
{\text \mjup} &  R \geq b
\end{cases}
\]

\noindent where $b$ is defined
below, and we use the weight $x=\left(R_{p}/\rearth-1\right)/3$ to
smoothly transition between $M_{\rm WM}$ and $M_L$ in their
overlapping regions of validity. $M_{\rm WM}$ is from \cite{Weiss2014} and is given by 
\[
\frac{M_{\rm WM}}{\mearth}=
\begin{cases}
0.615\left(0.717+\frac{R_{p}}{\rearth}\right)\left(\frac{R_{p}}{\rearth}\right)^3 &  R<1.5\rearth \\
2.69\left(\frac{R_{p}}{\rearth}\right)^{0.93} &  1.5\rearth\leq R<4\rearth
\end{cases}
\]

$M_L$ is given by $M_{p}/\mearth=\left(R_{p}/\rearth\right)^{2.06}$ from \cite{Lissauer2011}, and $M_M$ is given by 
\[
\log_{10}\left(\frac{M_{p}}{M_0}\right)=-w\left(\frac{b}{\left(R_{p}/\rearth\right)}-1\right)^{1/p}
\]

\noindent from \cite{Mordasini2012}, where
\[
\begin{tabular}{ c c c }
  & $a/{\rm AU} \leq 0.1$ & $a/{\rm AU} > 0.1$  \\ \hline $M_0[\mearth]$ & 1756.7 & 1308.7 \\
$b[\rearth]$ & 11.684 & 11.858 \\ $w$ & 1.646 & 1.635 \\ $p$ & 2.489 & 2.849 \\ 
\end{tabular}
\]

Planets with masses estimated in this manner are so noted in the
MASSREF field.

{\bf SEP} is a separation between the host star
and the planet in units of AU. For directly imaged and microlensing planets, the
literature usually reports the projected separation and we report the
value accordingly.  If SEP is not directly measured but the semi-major
axis ({\bf A}) is available, then we set SEP to A with its reference,
SEPREF, set to ``Set to A''. 

The parameter {\bf I} reports the orbital inclination of the planet. Previously,
only transiting systems had $i$ measured, but now some non-transiting
systems have dynamical or astrometric constraints on $i$ (for example,
the microlensing system OGLE-2006-BLG-109L by \citealt{Bennett2010}
and the imaged system, $\beta$ Pictoris by \citealt{Lagrange2009}).

{\bf BIGOM} is the longitude of the ascending node ($\Omega$) in units
of degrees. At the moment, only $\beta$ Pictoris $b$ has this quantity listed.

{\bf LAMBDA} is the projected spin-orbit misalignment ($\lambda$) in
the plane of the sky in units of degrees. At the moment, $\lambda$
values are only available for the transiting systems, which are
measured by either the Rossiter-McLaughlin effect
\citep[e.g.,][]{Winn2005} or planetary transits of star spots
\citep[e.g.,][]{Sanchis-Ojeda2012}. $\lambda$ is used because the true
spin-orbit angle, defined as the angle between the stellar spin axis
and the orbital axis, is normally not directly measurable. We follow
\cite{Fabrycky2009} who defined the quantity as follows: the $z$-axis
points toward the observer, the $x$-axis points along the intersection
of the sky plane and the planet's orbital plane with the ascending
node of the planet having $x<0$, and the $y$-axis completes a right
handed-triad. $\lambda$ is measured clockwise on the sky from the
$y$-axis to the projected stellar rotational axis. $\lambda$ is not a
newly added field to the EOD but we describe it here because we did
not previously report it consistently, since some literature report
the ``projected spin-orbit misalignment" as $\beta = -\lambda$
instead\footnote{In some cases, \citep[e.g.][]{Moutou2011}, other
  conventions entirely appear to be used.}. We have sorted these cases
out and made sure that our $\lambda$'s are following the definition of
\cite{Fabrycky2009}.

The {\bf MASSREF} and {\bf MASSURL} fields were used to report the
reference and the URL of stellar mass but now they are used for the planet's
mass; {\bf MSTARREF} and {\bf MSTARURL} are now used for the stellar
mass, {\bf MSTAR}.

\subsection{Transit Parameters}\label{sec:transit}

We have added two fields since \cite{Wright2011}, which are used for
KOI's incorporated from the Exoplanet Archive. 

{\bf DR} is the distance between the planet and the star during the
transit in stellar radii. This is one of the parameters that goes into
the transit light curve modeling (see e.g. \citealt{Batalha2013}).

{\bf RR} is the ratio of the planetary radius and the stellar
radius. If the transit depth (DEPTH) is not given, we calculate DEPTH
by taking square of RR, but we do not calculate RR from DEPTH.

\subsection{Secondary Eclipse}\label{sec:se}

\begin{figure}[!htb]
\centering
\includegraphics[width=\textwidth]{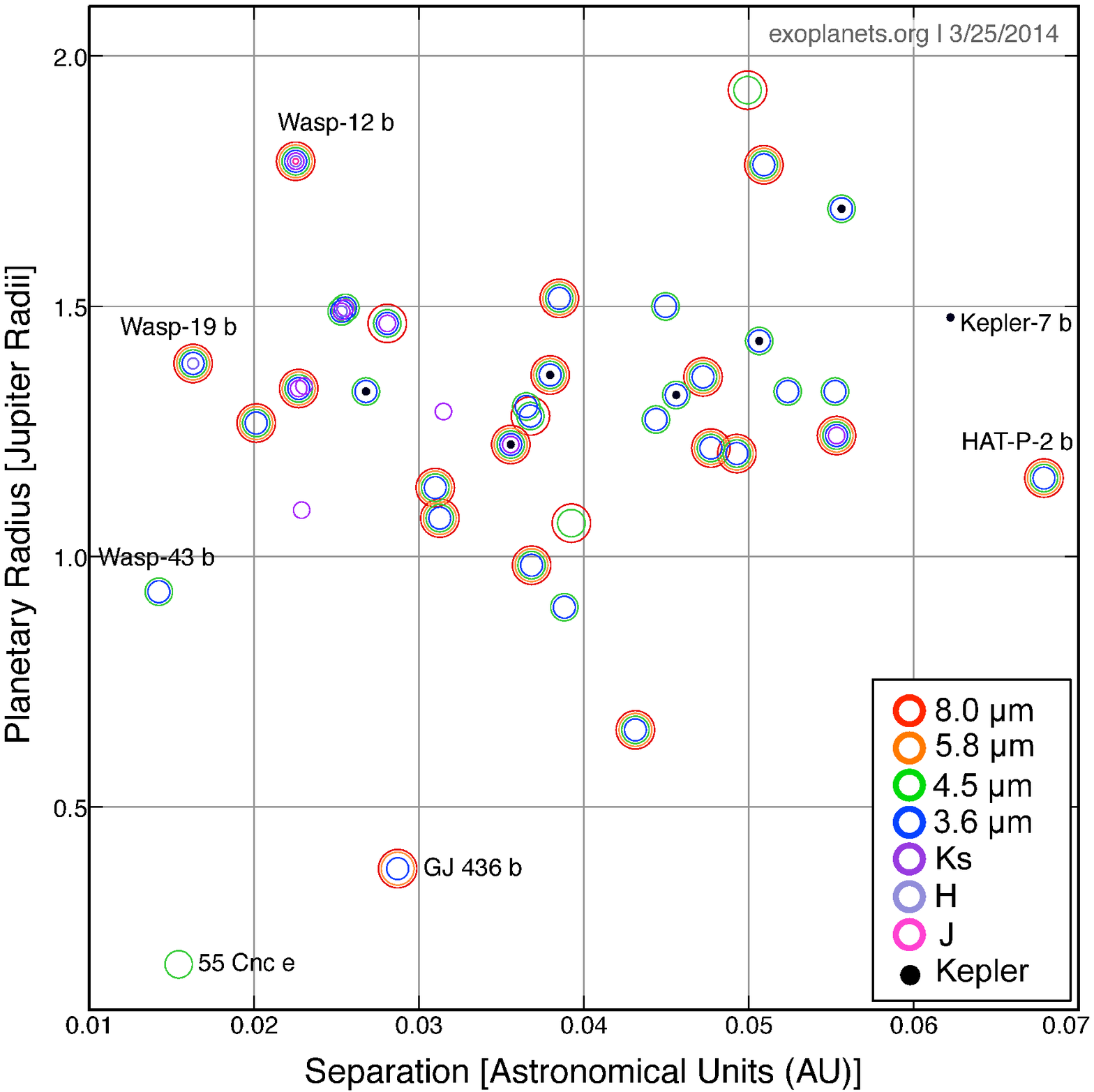}
\caption{An example plot that demonstrates the current secondary
  eclipse measurements. Each planet has a set of circles that stand
  for its available secondary eclipse measurements in the database,
  including \spitzer\ 8.0, 5.8, 4.5, \& 3.6 \micron, and ground-based
  $J$, $H$, and $K_{\rm s}$ bands. \kepler\ secondary eclipse measurements
  are labeled as solid black dots. A few representative planets are
  also annotated (a feature not offered by the EDE Plotter, but
  trivially implemented by other image or presentation software). The
  planet name will appear when the user points the cursor at the
  corresonding data point in the interactive plotting tools.}
\label{fig:se}
\end{figure}

We have added an entirely new set of fields for secondary
eclipse depths. Previously, due to the limits of instrumentation and limited
sample of exoplanets, there was not much secondary eclipse depth data
available. Since \cite{Wright2011}, as many different surveys have
been launched, secondary eclipse depths at multiple
bands have become available for many planets. Therefore, we added
this entire new set of fields (see Figure~\ref{fig:se}).

{\bf SEDEPTHX} whereas X stands for one of ``J", ``H", ``KS", ``KP",
``36", ``45", ``58", ``80": these fields are the secondary eclipse
depth measured in the corresponding wavelength band. SEDEPTHJ, H, and
KS are the depths measured in the 2MASS wavelength bands in the near
infrared, centered at 1.25~\micron, 1.65~\micron, and 2.15~\micron,
respectively. KP is the depth measured in the \kepler\ photometric
band which spans 400 to 865~nm. The other four represent the
\spitzer\ IRAC bands centered at 3.6~\micron, 4.5~\micron,
5.8~\micron, and 8.0~\micron.

The {\bf SE} Boolean indicates whether the system has secondary
eclipse depths measured in any band.  Truth indicates the
system has at least one secondary eclipse measurement in the EOD.

\subsection{Stellar Properties}\label{sec:stellarprop}

This set of parameters provides users the physical parameters of
the exoplanet host stars.  Most of the parameters are derived from the
stellar models in the literature and we provide the readers with the
references. We describe added fields since \cite{Wright2011} below.

{\bf GAMMA} ($\gamma$) is the radial velocity of the center of mass of
the planetary system with respect to the barycentric frame and is
reported in km/s.  It is typically known to far less precision than
$K$, because it requires an absolute measurement of velocity.

{\bf RSTAR} is the radius of the star in the solar units.

{\bf RHOSTAR} is the density of the star in g/cm$^3$.

\subsection{Stellar Magnitudes}\label{sec:stellarmag}

We provide the brightness of the host star measured in different
filters. We report optical magnitude and color ($V$ and $B-V$),
2MASS $J$, $H$, and $K_s$ magnitudes and, we have added {\bf KP} for
the \kepler\ photometric band. Note that the definitions of $V$ and $B$ in the literature vary, and are often Hipparcos magnitudes.

\subsection{Coordinates and Catalogs}\label{sec:coord}

We report the coordinates of the host star as well as the host star's
ID as designated by different catalogs.  We have added the {\bf DIST} field since
\cite{Wright2011}, which is the distance to the host star in the unit
of pc. If only parallax (PAR) is given, we calculate DIST based on PAR
and vice versa. If the literature does not report DIST or PAR, then we
obtain PAR from the Hipparcos dataset by \cite{van Leeuwen2009} for
stars with Hipparcos numbers. We na\"ively translate the 1-$\sigma$ errors in each into asymmetric
uncertainties in the other.

\subsection{Uncertainties}\label{sec:unc}

We have changed the way of reporting uncertainties, especially
asymmetric ones since \cite{Wright2011}. We now report
uncertainties for a certain field X as XUPPER, XLOWER and UX. XUPPER
and XLOWER store the upper and lower 1$\sigma$ uncertainties, and they
are different when the error bars are asymmetric. UX contains the
1$\sigma$ uncertainty, and it equals $({\rm XUPPER+XLOWER})/2$. Fields
that have XUPPER, XLOWER and UX are: all the orbital parameters, all
the transit parameters, all the secondary eclipse fields, and all the stellar
properties and DIST and PAR.

\subsection{References}\label{sec:ref}

The fields {\bf XREF} and {\bf XURL} store the peer-reviewed reference
and link to the publication for the data in field X. Previously, we 
reported only general references for certain sets of fields (e.g. FIRSTREF,
FIRSTURL, ORBREF, and ORBURL), but now we report the references for
most of the fields on an individual basis.

The following fields do not have associated REF or URL fields. Except
where we record these values directly from discovery papers, {\bf RA},
{\bf DEC}, and {\bf PAR} are usually from the \textit{Hipparcos}
re-reduction of \cite{van Leeuwen2009}; the $B-V$ ({\bf BMV}) values
are from the \textit{Hipparcos} catalog by \cite{Perryman1997};
stellar magnitudes {\bf J}, {\bf H}, and {\bf KS} (i.e.\ $K_{\rm s}$)
are from the 2MASS catalog \citep{Skrutskie2006} (as reported in
SIMBAD) and the {\bf KP} (i.e.\ $K_P$) magnitudes are from the
Exoplanet Archive \citep{Akeson2013}.

{\bf SHK} and {\bf RHK} represent the Mount Wilson S and
$R^\prime_{HK}$ chromospheric
emission measures.  These fields are poorly maintained fields and
unreferenced, at the moment. We plan to include proper reference in
the future. Most of the SHK and RHK values are from \cite{Butler2006},
and individual discovery papers. 

\subsection{Removed Fields}\label{sec:removed}

We removed 2 fields since \cite{Wright2011}:

The {\bf SPTYPE} field previously reported the stellar type of the host
star.  We have removed it since it can be difficult to verify with
peer-reviewed references, and the field was not well maintained.  We encourage users
to use TEFF and BMV instead.

The {\bf NSTEDID} field reported ID of host star in NStED, and now it is
replaced by EANAME, which is the name of the star as appears on the
NASA's Exoplanet Archive \citep{Akeson2013}.

\section{Inclusion of the Kepler Planet Candidates \\
  and Other Exoplanets on exoplanets.org}\label{sec:kepler}

Since \cite{Wright2011}, \kepler\ has discovered many exoplanet
candidates, and more planets have been robustly discovered by various
other methods, although not all have very well characterized orbits.  We have
expanded the EDE to include those planets for the convenience of
our users, and also to keep exoplanets.org complete in terms of confirmed
exoplanets.

We now provide an option for users to include the KOIs discovered by \kepler\ in the
EDE. Although KOIs are not counted as part of the EOD due to their
status of being ``candidates", we include them on our website so that
these KOIs are put into the context of all confirmed exoplanets. We
have used broad definitions for SEP and MASS so that they may be
compared to EOD planets, even though they share few strictly defined parameters.  The
KOIs listed in our table include three \kepler\ data releases (as of
July 2014), and exoplanets.org always provides the most up-to-date
release through a partnership with the NASA Exoplanet Archive.  

\begin{figure}[!htb]
\centering
\includegraphics[width=\textwidth]{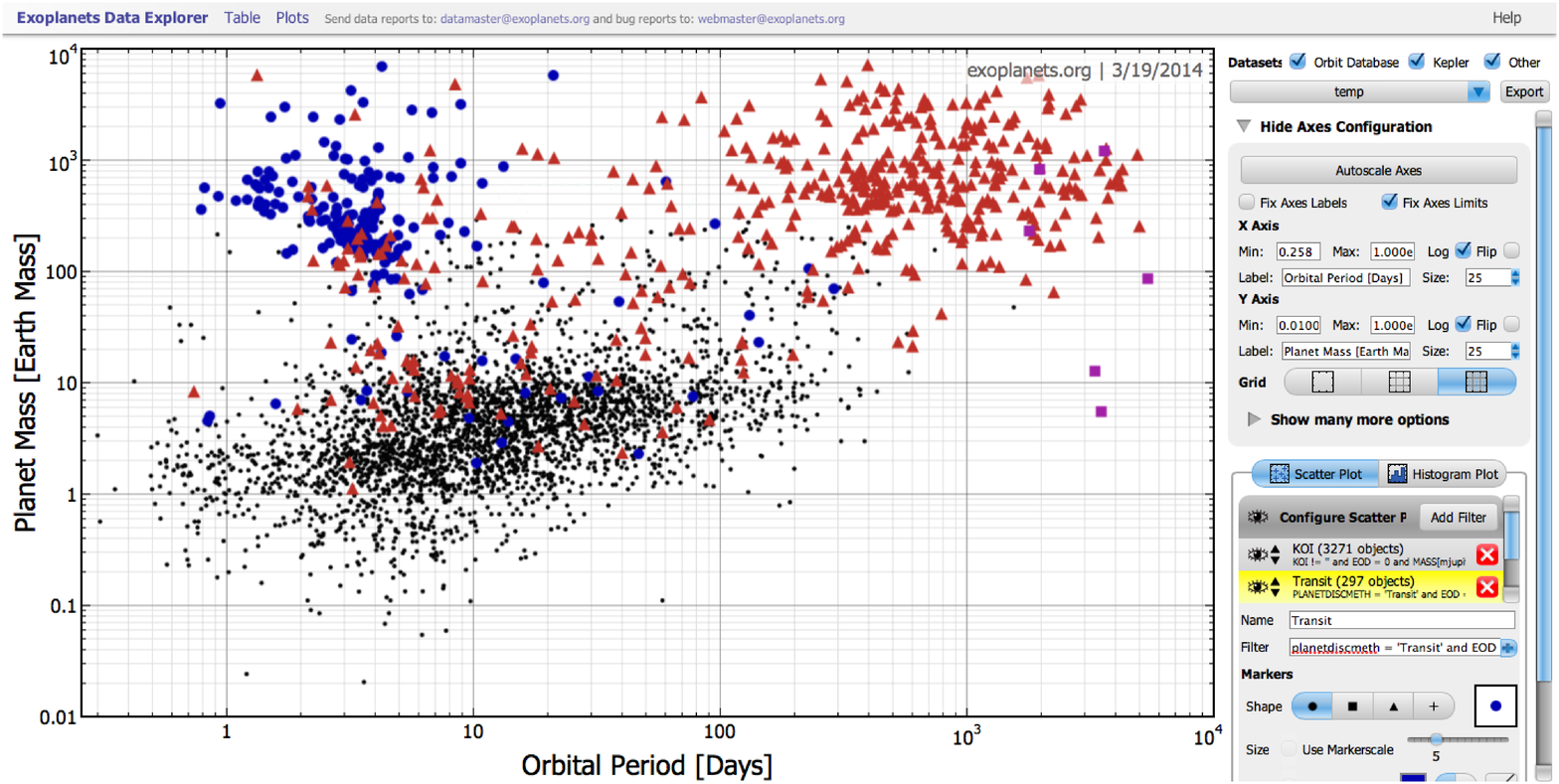}
\caption{
  An example plot produced by the interactive plotting tools of
  EDE on exoplanets.org. The interactive plotting interface is shown
  on the right, with the mass-period plot shown on the left. The
  $x$-axis is planet orbital period in days and $y$-axis is planet
  mass in Earth's mass. The exoplanet samples being plotted are the
  ones in the EOD, including transit planets (blue large dots), RV
  planets (red triangles), and timing planets (magenta squares). The
  KOIs are plotted as black small dots, and their masses are
  calculated using their radii based on mass-radius relations from the
  literature (see Section~\ref{sec:orbit}).  Note the
  apples-to-oranges comparison enabled by the hybrid MASS field on the y-axis, which blends estimated
  masses, minimum masses, and true masses on a single plot.}
\label{fig:koi}
\end{figure}

To display the KOIs in the EDE table, users can check the ``Kepler" option on
the upper left of the table (the default selection only includes ``Orbit
Database", i.e.~the EOD, and ``Other"). They can also be displayed on
plots with the interactive plotting tool of EDE (see
Figure~\ref{fig:koi} for an example plot).  All KOIs with a
disposition other than ``FALSE POSITIVE'' are included in the EDE.

We now also include robustly detected imaging, microlensing,
and timing planets.  Our criteria for a ``robust detection'' for these
planets are the
following:

\begin{itemize}
\item Imaging planets:
\begin{enumerate}
\item Planet-star mass ratio $q < 0.023$ ($< 24\ $\mjup\ for a solar
  mass star); and in general we require ($q+\sigma_q$) $< 0.023$,
  where $\sigma_q$ is the uncertainty in $q$.
\item SEP $<$ 100 AU $\times$ ($M_*/M_\odot$), where SEP is the
  semi-major axis a if a is known, and the projected separation
  otherwise.
\item Confidently detected: the detection is clearly of a real
  astrophysical source, and will unlikely later be found to have been
  spurious.
\item Confidently bound: object will clearly not
  be later found to be unbound or a chance alignment.

\end{enumerate}
\item Microlensing planets:\\
  We accept microlensing planets that appear in refereed journals as
  having unambiguously planetary masses (planet host mass ratio,
  $q<0.023$). 

\item Timing planets:\\
  Timing planets must have dynamically stable orbits and, ideally, show
  multiple complete orbits (e.g., see \citealt{Wittenmyer2012},
  \citealt{Horner2012}, and \citealt{Wittenmyer2013}). 

\end{itemize}
Some of the exoplanets that are primarily detected by these methods do
not have well-determined orbital parameters, and thus they are
categorized as ``Other Planets'' on exoplanets.org instead of being in
the EOD. These planets can be displayed by check the `Other' option
for the EDE table. As of July 2014, no microlensing or imaging
planets (except $\beta$ Pictoris) appear in the EOD, while several timing exoplanets have
well-determined orbital parameters and are included in the
EOD.

\section{Updates to the Website software}\label{sec:website}

The software stack powering the EOD remains largely unchanged from
\cite{Wright2011}. The backend server is now written in Ruby, instead
of Python, but the thin role it plays is largely unchanged.  The
JavaScript frontend that runs on the browser has recently been
optimized to support the larger datasets arising from the Kepler
missions.  In particular, the EOD table now only renders cells that
are visible on-screen. 
\begin{enumerate}
\item Plot parameters may be saved with browser cookies, so plots may
  be custom reproduced easily with updated data later. 
\item Tables may be generated and exported with custom fields
  computed. 
\item Plots may be exported with a variety of resolutions and aspects,
  including publication-quality vector graphics (SVG format).  
\end{enumerate}

\section{Summary and Future Work}\label{sec:summary}

We have had two major updates since \cite{Wright2011}. First, we
expanded the scope of the EOD. We removed the requirement for the host
star being normal and included the planets discovered not only by RV
and transit but also by various other methods (e.g. timing) in the
EOD. We also have added some new fields, most notably, we now report
the secondary eclipse depths. We removed obsolete fields and revised
some fields as necessary. We have included SEP and MASS with broad
definitions so all planets may be plotted on common axes.

Second, we included KOIs and other planets (the ones detected by
microlensing, imaging, timing, and in the future, astrometry but do
not have well characterized orbits) on exoplanets.org. Although they
are not included in the EOD, we reported them on our website since the
users might find them useful for purposes of statistical studies and
comparison.

In the future, we hope to include images of directly imaged planets
and light curves for transits and microlensing planets. We plan to
display RV curves whenever their RV data are available and include
better incorporation of Kepler parameters in the EOD (such as KEPID).

We also plan to report the upper limits (XUL) for planet minimum mass (MSINI),
eccentricity (ECC), and the radial velocity semi-amplitude (K) when
the literature reports them. For example, an upper limit for ECC is
reported when the orbit is consistent with a circular one, and the
upper limit for K is reported when there is only a null result from
radial velocities. Since not all authors define the upper limits in the same
way (e.g. 1$\sigma$ or some fixed percentile), we encourage the
users to refer to the original references for more information. We
will also include proper rendering of upper limits in the EDE Table,
proper upper limit symbols in the Plotter, and proper handling of
upper limits in filters and queries in both cases. 
 
Omissions and errors to the EOD can be corrected by sending us an
email at \url{datamaster@exoplanets.org}.

\acknowledgments

We acknowledge NSF grant AST-1211441 which supports exoplanets.org.
Parts of this work was supported by a NASA Keck PI Data Award, administered by
the NASA Exoplanet Science Institute.

We also thank NASA and NExScI for support via funds associated with
the award of Keck time for monitoring the orbits of long-period and
multi-planet systems.

SXW acknowledges support from the Penn State Astrobiology Research
Center, part of the NASA Astrobiology Institute (NNA09DA76A). The Center for
Exoplanets and Habitable Worlds is supported by the 
Pennsylvania State University, the Eberly College of Science, and the
Pennsylvania Space Grant Consortium.

This research has made use of
NASA's Astrophysics Data System Bibliographic Services (ADS) and the SIMBAD database,
operated at CDS, Strasbourg, France. 

We thank Jean Schneider and the exoplanet.eu team, Rachel Akeson
and the Exoplanet Archive team, and David Ciardi and the NStED team,
for their hard work on their exoplanet compendia and for their
occasional assistance and support.  Our parallel,
independent, and complementary efforts improve all of our resources.
We are especially grateful for the radial velocity archive compiled by
NStED and currently maintained by the Exoplanet Archive, which enables
many of the radial velocity plots at exoplanets.org.

We thank Katie Peek, the information graphics editor at Popular
Science magazine, for compiling the data to initially populate the
MONTH field of the EOD and sharing it with us.

We thank Jacob Bean for his referee report and suggestions to exoplanets.org.

We thank the many people who contributed bug reports and data
errors to \url{datamaster@exoplanets.org}, which continue to improve
the data quality of the EOD.


\clearpage

\begin{deluxetable}{llllc}
  \center
\tabletypesize{\scriptsize}
\tablewidth{0pt}
\tablecaption{Fields of Exoplanet Orbit Database\label{tab:par}}
\tablehead{
  \colhead{Field} & \colhead{Data Type} & \colhead{Meaning} &
  \colhead{Related Fields\tablenotemark{a}} & \colhead{Ref.\tablenotemark{b}}
}
\startdata
\sidehead{\textbf{Discovery and References}}
NAME\dotfill & String & Name of planet & \nodata & W11 \\
OTHERNAME\dotfill & String & Other commonly used name of star & \nodata & \ref{sec:disc} \\
DATE\dotfill & Integer & Year of publication of FIRSTREF & \nodata & W11 \\
MONTH\dotfill & Integer & Month of publication of FIRSTREF & \nodata & \ref{sec:disc} \\
PLANETDISCMETH\dotfill & String & Method of discovery of planet & \nodata & \ref{sec:disc} \\
STARDISCMETH\dotfill & String & Method of discovery of first planet in system & \nodata & \ref{sec:disc} \\
ORBREF\dotfill & String & Peer-reviewed origin of orbital parameters & ORBURL & W11 \\
FIRSTREF\dotfill & String & First peer-reviewed publication of
planetary orbit & FIRSTURL & W11 \\
JSNAME\dotfill & String & Name of host star used in the Extrasolar
Planets Encyclopaedia & EPEURL & W11 \\
ETDNAME\dotfill & String & Name of planet used in the Exoplanet
Transit Database & ETDURL & W11 \\
EANAME\dotfill & String & Name of planet used in the Exoplanet
Archive database & EAURL & \ref{sec:removed} \\
SIMBADNAME\dotfill & String & Valid SIMBAD name of host star (or
planet, if available) & SIMBADURL & W11 \\
KEPID\dotfill & Long Integer & The unique \kepler\ star identifier &
\nodata & \ref{sec:disc} \\
KOI\dotfill & Float & KOI object number & \nodata & \ref{sec:disc} \\
EOD\dotfill & Boolean & If true, planet is included in the EOD & \nodata & \ref{sec:disc} \\
MICROLENSING\dotfill & Boolean & If true, planet was detected via microlensing & \nodata & \ref{sec:disc} \\
IMAGING\dotfill & Boolean & If true, planet was detected via imaging & \nodata & \ref{sec:disc} \\
TIMING\dotfill & Boolean & If true, planet was detected via timing & \nodata & \ref{sec:disc} \\
ASTROMETRY\dotfill & Boolean & If true, planet was detected via astrometric motion & \nodata & \ref{sec:disc} \\
\sidehead{\textbf{Orbital Parameters}}
MSINI\dotfill & Float & Minimum mass (as calculated from the mass
function) in \mjup & -UL, -UPPER, etc. & W11 \\
MASS\dotfill & Float & Mass of planet in \mjup & -UPPER, etc. & \ref{sec:orbit} \\
A\dotfill & Float & Orbital semimajor axis in AU & -UPPER, etc. & W11 \\
SEP\dotfill & Float & Separation between host star and planet in AU & -UPPER, etc. & \ref{sec:orbit} \\
PER\dotfill & Double & Orbital period in days & -UPPER, etc. & W11 \\
K\dotfill & Float & Semiamplitude of stellar reflex motion in \mps &
-UL, -UPPER, etc. & W11 \\
ECC\dotfill & Float & Orbital eccentricity & -UL, -UPPER, etc. & W11 \\
I\dotfill & Float & Orbital inclination in degrees & -UPPER, etc. & \ref{sec:orbit} \\
OM\dotfill & Float & Argument of periastron in degrees & -UPPER, etc. & W11 \\
BIGOM\dotfill & Float & Longitude of ascending node in degrees & -UPPER, etc. & \ref{sec:orbit} \\
T0\dotfill & Double & Epoch of periastron in HJD\tablenotemark{c} & -UPPER, etc. & W11 \\
DVDT\dotfill & Float & Magnitude of linear trend in \mps\ day$^{-1}$ & -UPPER, etc. & W11 \\
LAMBDA\dotfill & Float & Projected spin-orbit misalignment in degrees
& -UPPER, etc. & \ref{sec:orbit} \\
TRANSIT\dotfill & Boolean & Is the planet known to transit? & -REF,-URL & W11 \\
\sidehead{\textbf{Transit Parameters}}
R\dotfill & Float & Radius of planet in Jupiter radii & -UPPER,
etc. & W11 \\
TT\dotfill & Float & Epoch of transit center in
HJD\tablenotemark{c} & -UPPER, etc. & W11 \\
T14\dotfill & Float & Time of transit from first to fourth contact in days & -UPPER, etc. & W11 \\
B\dotfill & Float & Impact parameter of transit & -UPPER, etc. & W11 \\
AR\dotfill & Float & $(a/R_*)$ & -UPPER, etc. & W11 \\
DEPTH\dotfill & Float & Transit depth, or $(R_p/R_*)^2$ when depth is
not given & -UPPER, etc. & W11 \\
DENSITY\dotfill & Float & Density of planet in g cc$^{-1}$ &
-UPPER, etc. & W11 \\
GRAVITY\dotfill & Float & $\log{g}$ (surface gravity) of planet in cgs unit &
-UPPER, etc. & W11 \\
DR\dotfill & Float & Distance during transit in stellar radii & -UPPER, etc. & \ref{sec:transit} \\
RR\dotfill & Float & $(R_p/R_*)$ & -UPPER, etc. & \ref{sec:transit} \\
\sidehead{\textbf{Orbital Fit Properties}}
CHI2\dotfill & Float & $\chi_{\nu}^2$ to orbital RV fit & \nodata & W11 \\
NOBS\dotfill & Integer & Number of observations used in fit & \nodata & W11 \\
RMS\dotfill & Float & Root-mean-square residuals to orbital RV fit &
\nodata & W11 \\
FREEZE\_ECC\dotfill & Boolean & Eccentricity frozen in fit? & \nodata & W11 \\
TREND\dotfill  & Boolean & Linear trend in fit? & \nodata & W11 \\
NCOMP\dotfill & Integer & Number of planetary companions known & \nodata & W11 \\
MULT\dotfill & Boolean & Multiple planets in system? & \nodata & W11 \\
COMP\dotfill & String & Component name of planet ($b$, $c$, etc.) & \nodata & W11 \\
\sidehead{\textbf{Secondary Eclipse Depth}}
SE\dotfill & Boolean & If true, at least one secondary eclipse has
been detected & -REF,-URL & \ref{sec:se} \\
SEDEPTHJ\dotfill & Float & Secondary eclipse depth in $J$ band & -UPPER, etc. & \ref{sec:se} \\
SEDEPTHH\dotfill & Float & Secondary eclipse depth in $H$ band & -UPPER, etc. & \ref{sec:se} \\
SEDEPTHKS\dotfill & Float & Secondary eclipse depth in $K_S$
band & -UPPER, etc. & \ref{sec:se} \\
SEDEPTHKP\dotfill & Float & Secondary eclipse depth in the
\kepler\ photometry band & -UPPER, etc. & \ref{sec:se} \\
SEDEPTH36\dotfill & Float & Secondary eclipse depth in
\spitzer\ IRAC1 3.6 \micron\ band & -UPPER, etc. & \ref{sec:se} \\
SEDEPTH45\dotfill & Float & Secondary eclipse depth in
\spitzer\ IRAC2 4.5 \micron\ band & -UPPER, etc. & \ref{sec:se} \\
SEDEPTH58\dotfill & Float & Secondary eclipse depth in
\spitzer\ IRAC3 5.8 \micron\ band & -UPPER, etc. & \ref{sec:se} \\
SEDEPTH80\dotfill & Float & Secondary eclipse depth in
\spitzer\ IRAC4 8.0 \micron\ band & -UPPER, etc. & \ref{sec:se} \\
%
\sidehead{\textbf{Stellar Properties}}
STAR\dotfill & String & Standard name for host star & \nodata & W11 \\
BINARY\dotfill & Boolean & Star known to be binary? & -REF, -URL & W11 \\
MSTAR\dotfill & Float & Mass of host star in solar mass & -UPPER, etc. & W11 \\
RSTAR\dotfill & Float & Radius of host star in solar radii & -UPPER, etc. & \ref{sec:stellarprop} \\
FE\dotfill & Float & Iron abundance (or metallicity) of star & -UPPER, etc. & W11 \\
TEFF\dotfill & Float & Effective temperature of host star in Kelvins & -UPPER, etc. & W11 \\
RHOSTAR\dotfill & Float & Density of host star in g/cm$^3$ & -UPPER, etc. & \ref{sec:stellarprop} \\
LOGG\dotfill & Float & Spectroscopic $\log{g}$ (surface gravity) of
host star in cgs unit & -UPPER, etc. & W11 \\
VSINI\dotfill & Float & Projected equatorial rotational velocity of
star in k\mps & -UPPER, etc. & W11 \\
GAMMA\dotfill & Float & Systemic radial velocity in k\mps & -UPPER, etc. & \ref{sec:stellarprop} \\
\sidehead{\textbf{Stellar Magnitudes}}
V\dotfill & Float & $V$ magnitude & -REF,-URL & W11 \\
BMV\dotfill & Float & $B-V$ color & \nodata & W11 \\
J\dotfill & Float & $J$ magnitude & \nodata & W11 \\
H\dotfill & Float & $H$ magnitude & \nodata & W11 \\
KS\dotfill & Float & $K_S$ magnitude & \nodata & W11 \\
SHK\dotfill & Float & Mount Wilson Ca {\sc{II}} $S$-value & \nodata & \ref{sec:ref} \\
RHK\dotfill & Float & Chromospheric activity of star as $\log{R'_{HK}}$ & \nodata & \ref{sec:ref} \\
KP\dotfill & Float & \kepler\ bandpass magnitude & \nodata & \ref{sec:stellarmag} \\
SPECREF\dotfill & String & Source of most of the spectroscopic parameters & SPECURL & W11 \\
\sidehead{\textbf{Coordinates and Catalogs}}
RA\dotfill & Double & J2000 right ascension in decimal hours & \nodata & W11 \\
RA\_STRING\dotfill & String & J2000 right ascension in sexagesimal string & \nodata  & W11 \\
DEC\dotfill & Double & J2000 declination in decimal degrees & \nodata & W11 \\
DEC\_STRING\dotfill & String & J2000 declination in sexagesimal string & \nodata & W11 \\
PAR\dotfill & Float & Parallax in mas & -UPPER,-LOWER,U- & W11 \\
DIST\dotfill & Float & Distance to host star based on parallax in parsecs & -UPPER, etc. & \ref{sec:coord} \\
HIPP\dotfill & Long Integer & \textit{Hipparcos} catalog number of
star & \nodata & W11 \\
HD\dotfill & Long Integer & Henry Draper number of sar & \nodata & W11 \\
GL\dotfill & Float & GJ or Gliese catalog number of star & \nodata & W11 \\
HR\dotfill & Integer & Bright Star Catalog number of star & \nodata & W11 \\
SAO\dotfill & Long Integer & SAO catalog number of star & \nodata & W11 \\
\enddata
\tablenotetext{a}{``-UPPER etc." means ``-UPPER, -LOWER, U-, -REF,
  -URL", where ``-" stands for the name of the field listed in the
  first column. These fields store the uncertainties and
  references. See Section~\ref{sec:unc} and \ref{sec:ref} for more
  details. ``-UL" stands for upper limits for the fields (see
  Section~\ref{sec:unc} for details).}
\tablenotetext{b}{Fields that remain unchanged are referenced to
  \cite{Wright2011}, abbreviated as ``W11". Fields that are new or
  revised since \cite{Wright2011} are referenced to the relevant
  section in this paper.}
\tablenotetext{c}{As noted in \cite{Wright2011}, we do not report the
  epoches of periastron, transit center, and secondary eclipse center
  all consistently as HJD. We simply record what the original articles
  report, which are in varies formats (JD, BJD, HJD etc.).}
\end{deluxetable}

\end{CJK*}


\begin{thebibliography}

\bibitem[Akeson et al.(2013)]{Akeson2013} Akeson, R.~L., et al.\ 2013, 
\pasp, 125, 989 

\bibitem[Batalha et al.(2013)]{Batalha2013} Batalha, N.~M., et al.\ 2013, 
\apjs, 204, 24 

\bibitem[Bennett et al.(2010)]{Bennett2010} Bennett, D.~P., et al.\ 2010, 
\apj, 713, 837 

\bibitem[Borucki et al.(2010)]{Borucki2010} Borucki, W.~J., et al.\ 2010, 
Science, 327, 977 

\bibitem[Butler et al.(2002)]{Butler2002} Butler, R.~P., et al.\ 2002, 
\apj, 578, 565 

\bibitem[Butler et al.(2006)]{Butler2006} Butler, R.~P., et al.\ 2006, 
\apj, 646, 505 

\bibitem[Chauvin et al.(2012)]{Chauvin2012} Chauvin, G., et al.\ 2012, 
\aap, 542, A41 



\bibitem[Dawson \& Murray-Clay(2013)]{Dawson2013} Dawson, R.~I., \&
  Murray-Clay, R.~A.\ 2013, \apjl, 767, L24  

\bibitem[Fabrycky \& Winn(2009)]{Fabrycky2009} Fabrycky, D.~C., \&
  Winn, J.~N.\ 2009, \apj, 696, 1230 
  
\bibitem[Green et al.(2012)]{Green2012} Green, J., et al.\ 2012, 
arXiv:1208.4012   

\bibitem[Horner et al.(2012)]{Horner2012} Horner, J., Wittenmyer, R.~A., 
Hinse, T.~C., \& Tinney, C.~G.\ 2012, \mnras, 425, 749 

\bibitem[Howard(2013)]{Howard2013} Howard, A.~W.\ 2013, Science, 340,
  572 

  
\bibitem[Kane(2014)]{Kane2014} Kane, S.~R.\ 2014, \apj, 782, 111 
  
\bibitem[Kipping(2013)]{Kipping2013} Kipping, D.~M.\ 2013, \mnras,
  434, L51 

\bibitem[Lagrange et al.(2009)]{Lagrange2009} Lagrange, A.-M., Gratadour, D., Chauvin, G., et al.\ 2009, \aap, 493, L21 

\bibitem[Lissauer et al.(2011)]{Lissauer2011} Lissauer, J.~J., et al.\ 
2011, \apjs, 197, 8 

\bibitem[Macintosh et al.(2014)]{Macintosh2014} Macintosh, B., Gemini 
Planet Imager instrument team, Planet Imager Exoplanet Survey, G., 
\& Observatory, G.\ 2014, American Astronomical Society Meeting
Abstracts, 223, \#229.02  


\bibitem[Mahadevan et al.(2012)]{Mahadevan2012} Mahadevan, S., 
Ramsey, L., Bender, C., et al.\ 2012, \procspie, 8446, 

\bibitem[Mordasini et al.(2012)]{Mordasini2012} Mordasini, C., Alibert, Y., 
Georgy, C., Dittkrist, K.-M., Klahr, H., \& Henning, T.\ 2012, \aap, 547, A112

\bibitem[Moutou et al.(2011)]{Moutou2011} Moutou, C., D{\'{\i}}az, R.~F., Udry, S., et al.\ 2011, \aap, 533, A113  

\bibitem[Nielsen et al.(2014)]{Nielsen2014} Nielsen, E.~L., et al.\ 2014, 
arXiv:1403.7195 
  
\bibitem[Perryman \& ESA(1997)]{Perryman1997} Perryman, M.~A.~C., \&
  ESA 1997, ESA Special Publication, 1200 

\bibitem[Ricker(2014)]{Ricker2014} Ricker, G.~R.\ 2014, Journal of the 
American Association of Variable Star Observers (JAAVSO), 42, 234 


\bibitem[Sanchis-Ojeda et al.(2012)]{Sanchis-Ojeda2012} Sanchis-Ojeda, R., 
et al.\ 2012, \nat, 487, 449 

\bibitem[Schneider et al.(2011)]{Schneider2011} Schneider, J., Dedieu, C., 
Le Sidaner, P., Savalle, R., \& Zolotukhin, I.\ 2011, \aap, 532, A79

\bibitem[Skrutskie et al.(2006)]{Skrutskie2006} Skrutskie, M.~F., et al.\ 
2006, \aj, 131, 1163 


\bibitem[van Leeuwen(2009)]{van Leeuwen2009} van Leeuwen, F.\ 2009, \aap, 
497, 209 

\bibitem[Weiss \& Marcy(2014)]{Weiss2014} Weiss, L.~M., \& Marcy,
  G.~W.\ 2014, \apjl, 783, L6  

\bibitem[Winn et al.(2005)]{Winn2005} Winn, J.~N., et al.\ 2005, \apj, 631, 
1215 

\bibitem[Wittenmyer et al.(2012)]{Wittenmyer2012} Wittenmyer, R.~A., 
Horner, J., Marshall, J.~P., Butters, O.~W., 
\& Tinney, C.~G.\ 2012, \mnras, 419, 3258 

\bibitem[Wittenmyer et al.(2013)]{Wittenmyer2013} Wittenmyer, R.~A., 
Horner, J., \& Marshall, J.~P.\ 2013, \mnras, 431, 2150 

\bibitem[Wright et al.(2011)]{Wright2011} Wright, J.~T., et al.\ 2011,
  \pasp, 123, 412 

\bibitem[Wright \& Gaudi(2013)]{Wright2013} Wright, J.~T., \& Gaudi, B.~S.\ 2013,
  Planets, Stars and Stellar Systems.~Volume 3: Solar and Stellar
  Planetary Systems, 489 

\bibitem[Wright et al.(2014)]{Wright2014} Wright, J., et al.\ 2014, 
American Astronomical Society Meeting Abstracts, 223, \#148.31

\end{thebibliography}
\end{document}